\documentclass[aps,prc,preprint,amsmath,amssymb,showpacs,preprintnumbers,superscriptaddress]{revtex4-1}
\usepackage{CJK}
\usepackage{graphicx}
\usepackage{dcolumn}
\usepackage{bm}
\usepackage{mathrsfs}   
\usepackage{color,xcolor}  
\usepackage{multirow}  
\usepackage{booktabs} 

\usepackage{hyperref}

\newcommand{\la}{\langle}
\newcommand{\ra}{\rangle}

\newcommand{\bbm}{\begin{bmatrix}}
\newcommand{\ebm}{\end{bmatrix}}
\newcommand{\bBm}{\begin{Bmatrix}}
\newcommand{\eBm}{\end{Bmatrix}}
\newcommand{\bpm}{\begin{pmatrix}}
\newcommand{\epm}{\end{pmatrix}}

\begin{document}


\title{Neutron-proton effective mass splitting in neutron-rich matter}

\author{Sibo Wang}
\affiliation{Department of Physics, Chongqing University, Chongqing 401331, China}

\author{Hui Tong}
\affiliation{Helmholtz-Institut f$\ddot{u}$r Strahlen- und Kernphysik and Bethe Center for Theoretical Physics, Universit$\ddot{a}$t Bonn, D-53115 Bonn, Germany}

\author{Qiang Zhao}
\affiliation{Center for Exotic Nuclear Studies, Institute for Basic Science, \\
Daejeon 34126, Korea}

\author{Chencan Wang}
\affiliation{School of Physics and Astronomy, Sun Yat-Sen University, Zhuhai 519082, China}

\author{Peter Ring}
\affiliation{Department of Physik, Technische Universit\"{a}t M\"{u}nchen, D-85747 Garching, Germany}

\author{Jie Meng}
\email{mengj@pku.edu.cn}
\affiliation{State Key Laboratory of Nuclear Physics and Technology, School of Physics, \\
Peking University, Beijing 100871, China}
\affiliation{Yukawa Institute for Theoretical Physics, Kyoto University, \\
Kyoto 606-8502, Japan}

\date{\today}

\begin{abstract}

Nucleon effective masses in neutron-rich matter are studied with the relativistic Brueckner-Hartree-Fock (RBHF) theory in the full Dirac space.
The neutron and proton effective masses for symmetric nuclear matter are 0.80 times rest mass, which agrees well with the empirical values.
In neutron-rich matter, the effective mass of the neutron is found to be larger than that of the proton, and the neutron-proton effective mass splittings at the empirical saturation density are predicted as $0.187\alpha$ with $\alpha$ being the isospin asymmetry parameter.
The result is compared to other \emph{ab\ initio} calculations and is consistent with the constraints from the nuclear reaction and structure measurements, such as the nucleon-nucleus scattering, the giant resonances of $^{208}$Pb, and the Hugenholtz--Van Hove theorem with systematics of nuclear symmetry energy and its slope.
The predictions of the neutron-proton effective mass splitting from the RBHF theory in the full Dirac space might be helpful to constrain the isovector parameters in phenomenological density functionals.

\end{abstract}

\maketitle




Nucleon effective mass is a common concept used in the literature to characterize the momentum or energy dependence of the single-nucleon potential in the strongly interacting medium \cite{Jeukenne1976_PR25-83,Jaminon_PRC-1989-40-354}.
This quantity is essential for understanding many interesting phenomena in both nuclear physics and astrophysics
\cite{2018-LiBA-Prog.Phys.Nucl.Phys.}, including
the mean free path in a nucleus \cite{1981-Negele-PhysRevLett.47.71},
the nuclear liquid-gas phase transition \cite{2015-XuJ-PhysRevC.91.014611},
the in-medium nucleon-nucleon $(NN)$ cross-sections \cite{2005-LiBA-PhysRevC.72.064611},
and the stellar neutrino emission \cite{2014-Baldo-PhysRevC.89.048801}.
One of the hottest topics is the sign and the magnitude of the neutron-proton effective mass splitting $(m^*_n-m^*_p)/m$ in neutron-rich matter \cite{2007-Meissner-EPJA-31, 2007-Meissner-EPJA-32,ChenLW-2007-PhysRevC.76.054316}.
Due to the poor knowledge about the isovector channel of the in-medium $NN$ interaction and the lack of reliable experimental probes of the neutron-proton effective mass splitting, the quantity $(m^*_n-m^*_p)/m$ has not been well constrained.

In recent decades, large efforts have been devoted to constraining the neutron-proton effective mass splitting through various nuclear reaction and structure experiments, such as the nucleon-nucleus scattering \cite{2010-XuChang-PhysRevC.82.054607,LiXH-2015-Phys.Lett.B743.408}, the isovector giant dipole resonances \cite{2016-Zhang-PhysRevC.93.034335,2020-Xu-Phys.lett.B810.135820, 2017-Kong-PhysRevC.95.034324, 2021-Zhang-Chin.Phys.C}, and the heavy-ion collisions \cite{2013-Xie-PhysRevC.88.061601, 2015-Kong-PhysRevC.91.047601, 2016-Su-PhysRevC.94.034619, 2019-Morfouace-Phys.Lett.B799.135045}.
The neutron-proton effective mass splitting can also be explicitly related to the nuclear symmetry energy and its slope \cite{2005-Fritsch-Nucl.Phys.A750.259,2010-XuChang-PhysRevC.82.054607,2011-XuChang-Nucl.Phys.A865,ChenR-2012-PhysRevC.85.024305} according to the Hugenholtz--Van Hove (HVH) theorem \cite{Hugenholtz-1958-Physica}.
The constraints on the magnitude of $(m^*_n-m^*_p)/m$ from those methods vary from $-0.14\alpha$ to $0.56\alpha$ with $\alpha$ the isospin asymmetry, exhibiting a significant model dependence.
While many studies have found circumstantial evidences that the neutron-proton effective mass splitting is positive in neutron-rich matter, there are still several studies that favor the opposite.
Therefore, theoretical investigations on the neutron-proton effective mass splitting starting from first principles are particularly necessary.

Starting from realistic $NN$ interactions obtained from $NN$ scattering, numerous nuclear \emph{ab initio} many-body methods have been performed to study the neutron-proton effective mass, including the Brueckner-Hartree-Fock (BHF) theory \cite{1991-Baldo-PhysRevC.43.2605, 1999-Zuo-PhysRevC.60.024605, 2005-Zuo-PhysRevC.72.014005, 2014-Baldo-PhysRevC.89.048801, 2020-ShangXL-PhysRevC.101.065801, 2021-ShangXL-PhysRevC.103.034316},
the relativistic Brueckner-Hartree-Fock (RBHF) theory with the projection method \cite{VanDalen-2005-Phys.Rev.Lett.95.022302, VanDalen-2005-PRC72.065803, 2010-Dalen-PhysRevC.82.014319},
the self-consistent Green's function (SCGF) approach \cite{2004-Hassaneen-PhysRevC.70.054308},
and the many-body perturbation theory (MBPT) \cite{2016-Holt-PhysRevC.93.064603, Whitehead-2021-PhysRevLett.127.182502}.
Compared to the aforementioned experimental constraints, all the \emph{ab initio} studies predict a
positive effective mass splitting.
Among \emph{ab initio} methods, the RBHF theory is one of the most important methods in the relativistic framework, which contains essential three-body force effects and relativistic effects self-consistently \cite{Brown1987_CNPP17-39, Sammarruca-2012-PhysRevC.86.054317}.
The empirical saturation properties of symmetric nuclear matter (SNM) \cite{Brockmann1990_PRC42-1965} and the ground state properties of finite nuclei \cite{SHEN-SH2017_PRC96-014316} can be described satisfactorily within the RBHF theory.


Recently, by considering the nucleon positive-energy states (PESs) and negative-energy states (NESs) simultaneously, the RBHF equations are solved self-consistently in the full Dirac space \cite{WANG-SB2021_PRC103-054319,2022-Wang-SIBO-PhysRevC.106.L021305,Tong_2022-AstrophysicsJ930.137,WANG-SB2022_PhysRevC.105.054309,2023-QuXY-SciChina}.
The single-particle potentials are determined uniquely, which avoids the ambiguities caused by the neglect of NESs in the projection method \cite{Gross-Boelting1999_NPA648-105,VanDalen-2004-NPA744.227} and momentum-independence approximation method \cite{Brockmann1990_PRC42-1965,2003-Alonso-PhysRevC.67.054301,2023-QPP-Phys.Rev.D}.
The saturation properties of the SNM are reproduced in good agreement with the empirical values \cite{WANG-SB2021_PRC103-054319}.
In particular, the applications of the RBHF theory in the full Dirac space to asymmetric nuclear matter (ANM) have clarified the long-standing controversy about the isospin dependence of the Dirac mass in RBHF calculations \cite{2022-Wang-SIBO-PhysRevC.106.L021305}.
In this Letter, we use the RBHF theory in the full Dirac space to study the neutron-proton effective mass splitting in neutron-rich matter.


In the RBHF theory, the nucleon inside the nuclear medium is viewed as a dressed particle due to its interaction with the surrounding nucleons.
The single-particle motion of a nucleon in the nuclear medium is depicted by the Dirac equation, where the in-medium effects are contained in the single-particle potential operator $\mathcal{U}_\tau$.
It can be decomposed into the Lorentz form with three components \cite{Serot1986_ANP16-1}
\begin{equation}\label{eq:SPP}
  \mathcal{U}_\tau(\bm{p}) = U_{S,\tau}(p)+ \gamma^0U_{0,\tau}(p) + \bm{\gamma\cdot\hat{p}}U_{V,\tau}(p),
\end{equation}
where $U_{S,\tau}(p)$, $U_{0,\tau}(p)$, and $U_{V,\tau}(p)$ are the scalar potential, timelike, and spacelike parts of the vector potential, respectively. $\hat{\bm{p}}=\bm{p}/|\bm{p}|$ is the unit vector parallel to the momentum $\bm{p}$.

The $NN$ interaction of the dressed nucleon is also modified from its vacuum counterpart to the effective one.
In the RBHF calculation, the effective $NN$ interaction, named $G$ matrix, is obtained as the solution of the Thompson equation \cite{Brockmann1990_PRC42-1965}, which describes the two-nucleon scattering in the nuclear medium
\begin{equation}\label{eq:ThomEqu}
  \begin{split}
  G_{\tau\tau'}(\bm{q}',\bm{q}|\bm{P},W)
  =&\ V_{\tau\tau'}(\bm{q}',\bm{q}|\bm{P})
  + \int \frac{d^3k}{(2\pi)^3}
  V_{\tau\tau'}(\bm{q}',\bm{k}|\bm{P}) \\
    & \times
    \frac{Q_{\tau\tau'}(\bm{k},\bm{P})}{W-E_{\bm{P}+\bm{k},\tau}-E_{\bm{P}-\bm{k},\tau'} + i\epsilon}  G_{\tau\tau'}(\bm{k},\bm{q}|\bm{P},W),
  \end{split}
\end{equation}
where $\tau\tau'=nn$, $pp$, or $np$. Here $\bm{P}=\frac{1}{2}({\bm k}_1+{\bm k}_2)$ is the center-of-mass momentum, and $\bm{k}=\frac{1}{2}({\bm k}_1-{\bm k}_2)$ is the relative momentum of the two interacting nucleons with momenta ${\bm k}_1$ and ${\bm k}_2$.
The initial, intermediate, and final relative momenta of the two nucleons scattering in nuclear matter are $\bm{q}, \bm{k}$ and $\bm{q}'$, respectively.
The starting energy $W$ denotes the sum of the single-particle energies of two nucleons in the initial states~\cite{WANG-SB2021_PRC103-054319}.
The Pauli operator $Q_{\tau\tau'}(\bm{k},\bm{P})$ prohibits the nucleons from scattering to the occupied states.

One of the most important procedures in RBHF calculations is extracting the scalar and vector components of the single-particle potentials self-consistently from the $G$ matrix.
Without considering the NESs in the Dirac space, the extraction is usually done with the momentum-independence approximation or the projection method.
The momentum-independence approximation method~\cite{Brockmann1990_PRC42-1965,LI-GQ1992_PRC45-2782,2003-Alonso-PhysRevC.67.054301,TONG-H2018_PRC98-054302,WangCC-2020-JPG47.105108} assumes that the single-particle potentials are momentum independent and the spacelike part of the vector potential is negligible.
The scalar potential and the timelike part of the vector potential are then extracted directly from the single-particle potential energies at two selected momenta.
In the projection method~\cite{1987-Horowitz-Nucl.Phys.A,1997-Sehn-Phys.Rev.C,1998-Fuchs-Phys.Rev.C,1999-Gross-Boelting-Nucl.Phys.A,VanDalen-2004-NPA744.227,VanDalen-2007-Eur.Phys.J.A31-29}, the $G$ matrix elements are projected onto a complete set of five Lorentz invariant amplitudes~\cite{1987-Horowitz-Nucl.Phys.A}, from which the single-particle potentials are calculated.

These two methods suffer from the uncertainties of single-particle potentials due to the neglect of the NESs.
This can be avoided by implementing the RBHF calculation in the full Dirac space \cite{Anastasio1981_PRC23-2273}.
By solving the Thompson equation \eqref{eq:ThomEqu} in the full Dirac space, one obtains the $G$ matrix elements between both PESs and NESs, such as $\la \bar{u} \bar{u}|G|uu\ra$, $\la \bar{v} \bar{u}|G|uu\ra$, and $\la \bar{v} \bar{u}|G|vu\ra$, where $u$'s and $v$'s denote the PESs and NESs respectively.
Then the matrix elements of the single-particle potential operator $\la \bar{u}|\mathcal{U}_\tau(\bm{p})|u\ra$, $\la \bar{v}|\mathcal{U}_\tau(\bm{p})|u\ra$, and $\la \bar{v}|\mathcal{U}_\tau(\bm{p})|v\ra$ are calculated by summing up the $G$ matrix with all the nucleons inside the Fermi sea in the Hartree-Fock approximation.
From these matrix elements, the scalar and vector components of the single-particle potentials can be extracted uniquely \cite{WANG-SB2021_PRC103-054319}.

In relativistic models, the definition of the effective mass is often confused with that of the Dirac mass.
The former parametrizes the momentum and energy dependence of the single-particle potential, which is also known as the nonrelativistic effective mass.
The latter is defined through the scalar part of the nucleon self-energy in the Dirac equation.
To calculate the nonrelativistic effective mass, a quantity of practical importance is the so-called Schr\"odinger-equivalent potential $U_{\text{S.e.},\tau}$, which is extracted by reducing the Dirac equation to a Schr\"odinger-like equation and defined through
\begin{equation}\label{eq:Use}
  U_{\text{S.e.},\tau} = \frac{1}{2M}\left( U_{S,\tau}^2 - U^2_{0,\tau} + U^2_{V,\tau}
    + 2\varepsilon_\tau U_{0,\tau} + 2MU_{S,\tau} + 2MU_{0,\tau} + 2pU_{V,\tau} \right),
\end{equation}
where $\varepsilon_\tau$ is the single-particle energy of a nucleon without the rest mass $M$.
Starting from the Schr\"odinger-equivalent potential, the nonrelativistic effective mass can be calculated as \cite{VanDalen-2005-Phys.Rev.Lett.95.022302,VanDalen-2005-PRC72.065803}
\begin{equation}\label{eq:NRmass}
  M^*_{NR,\tau}(p) = \left[ \frac{1}{M} + \frac{1}{p}\frac{d}{dp}\mathrm{Re}\,U_{\text{S.e.},\tau}(p)\right]^{-1}.
\end{equation}
In practice, one focuses on its value at the Fermi momentum $M^*_{NR,\tau}(k^\tau_F)$, which is comparable to the effective mass $m^*_\tau$ derived from the analysis of nuclear structure and reaction experiments in the nonrelativistic framework \cite{2018-LiBA-Prog.Phys.Nucl.Phys.}.



\begin{figure}[htbp]
  \centering
  \includegraphics[width=16.0cm]{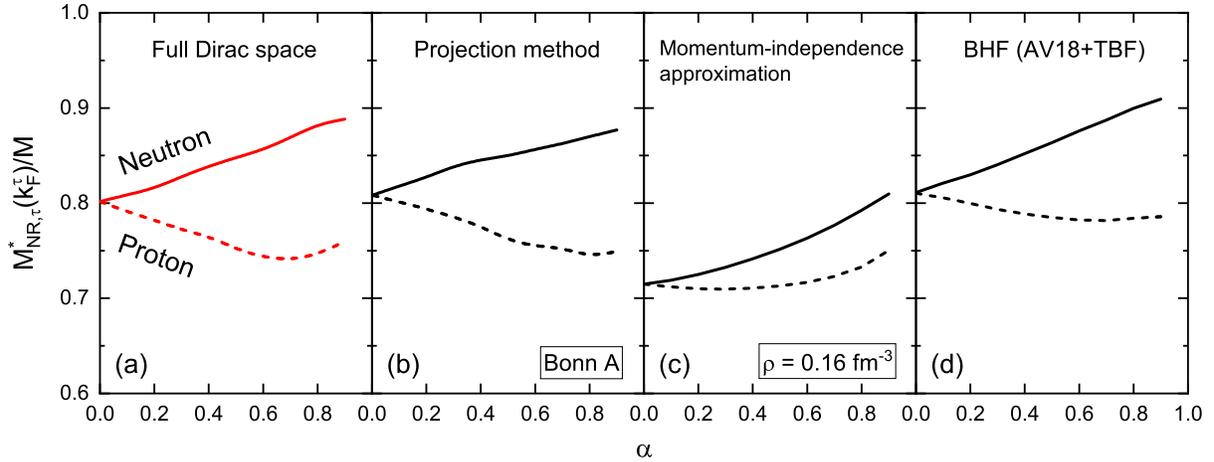}
  \caption{(Color online) The effective masses at Fermi momentum $M^*_{NR,\tau}(k_F^\tau)$ as functions of the asymmetry parameter $\alpha$ at the density $\rho=0.16\ \text{fm}^{-3}$ calculated by the RBHF theory in the full Dirac space \cite{2022-Wang-SIBO-PhysRevC.106.L021305} (a), in comparison with the results obtained by the RBHF theory with the projection method \cite{1999-Gross-Boelting-Nucl.Phys.A} (b) and the momentum-independence approximation method \cite{Brockmann1990_PRC42-1965} (c). The results from the BHF theory are also shown for comparison \cite{2021-ShangXL-PhysRevC.103.034316} (d).}
  \label{label-fig1}
\end{figure}

In panel (a) of Fig.~\ref{label-fig1}, the effective masses at Fermi momentum $M^*_{NR,\tau}(k^\tau_F)$ calculated by the RBHF theory in the full Dirac space with the Bonn A potential \cite{Machleidt1989_ANP19-189} are depicted as functions of the asymmetry parameter $\alpha=(\rho_n-\rho_p)/(\rho_n+\rho_p)$, where the density is chosen at the empirical nuclear saturation density 0.16\ $\text{fm}^{-3}$.
The results are compared to that obtained by the RBHF theory with the projection method [panel (b)], the RBHF theory with the momentum-independence approximation method [panel (c)], and the nonrelativistic BHF theory by using Argonne $v_{18}$ (AV18) potential with three-body forces \cite{2021-ShangXL-PhysRevC.103.034316} [panel (d)].

For SNM with $\alpha=0$, the effective masses for neutron and proton calculated in the full Dirac space in the unit of rest mass have the same value of 0.80, which agrees well with the empirical values of isoscalar effective mass $0.8\pm0.1$ constrained from the isoscalar giant quadrupole resonance (ISGQR) in $^{208}$Pb \cite{2018-LiBA-Prog.Phys.Nucl.Phys.}.
The results calculated by the projection method and the BHF theory are 0.81, which are close to the RBHF result in the full Dirac space.
However, the effective mass for SNM is underestimated by the momentum-independence approximation method.

With the increase of the asymmetry parameter, the effective mass for neutron in the full Dirac space increases steadily.
Similar tendencies are found for the projection method and the BHF theory.
The effective mass for the neutron calculated with the momentum-independence approximation method is also increasing, but the amplitudes are smaller than other calculations.
In the full Dirac space, with the increasing $\alpha$, the effective mass for the proton first decreases for $\alpha\leq 0.6$ and then increases for larger isospin asymmetry, showing an up-bending behavior.
The projection method shows qualitatively and quantitatively consistent results as those in the full Dirac space.
This implies that, for nuclear matter with small isospin asymmetry, the approximation schemes and techniques used in the projection method are reliable to determine the isospin dependent effective mass.
As for the momentum-independence approximation method, the isospin dependence of the proton effective mass shows apparent deviations in comparison to the results in the full Dirac space and the projection method.
This again verifies the conclusion found in previous studies \cite{Ulrych1997_PRC56-1788,2022-Wang-SIBO-PhysRevC.106.L021305}, that the momentum-independence approximation method fails to determine the correct behavior of the isospin dependence of the single-particle potentials in ANM.
For the BHF theory, with the increasing isospin asymmetry, a weak up-bending behavior of the effective mass for proton at large asymmetry is also found.


\begin{figure}[htbp]
  \centering
  \includegraphics[width=8.0cm]{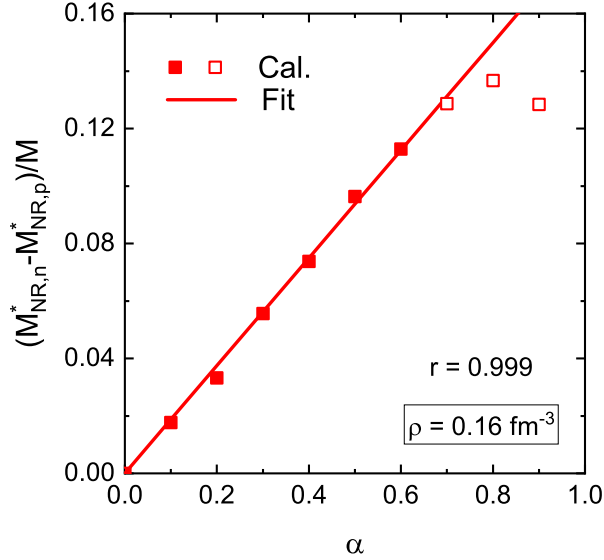}
  \caption{(Color online) The neutron-proton effective mass splitting as a function of the asymmetry parameter $\alpha$ at the density $\rho=0.16\ \text{fm}^{-3}$ calculated by the RBHF theory in the full Dirac space (solid and empty squares). The line represents the linear fitting for data in solid squares.}
  \label{label-fig2}
\end{figure}

By expanding the Schr\"odinger-equivalent potential as a power series of the asymmetry parameter $\alpha$ to the first order \cite{2010-XuChang-PhysRevC.82.054607,LiBA-2013-Phys.Lett.B727.276}, the neutron and proton effective masses are found proportional to $\alpha$.
As a consequence, the effective mass splitting $(M^*_{NR,n}-M^*_{NR,p})/M$ is also linearly dependent on $\alpha$.
In Fig.~\ref{label-fig2}, the splitting $(M^*_{NR,n}-M^*_{NR,p})/M$ calculated by the RBHF theory in the full Dirac space is shown as a function of isospin asymmetry.
It is found that the linear relation is valid in the range of $0\leq \alpha \leq 0.6$.
For larger asymmetry, due to the up-bending tendency of the effective mass for proton as shown in panel (a) in Fig.~\ref{label-fig1}, the effective mass splitting shows a non-monotonical behavior.

To extract the magnitude of the neutron-proton effective mass splitting, the splitting $(M^*_{NR,n}-M^*_{NR,p})/M$ in the range $0\leq\alpha\leq0.6$ is fitted with a linear function, where the Pearson's correlation coefficient $r$ is $0.999$, which confirms the validity of the assumption of a linear relation.
Invoking the fitting procedure, the magnitude of the neutron-proton effective mass is predicted as $(M^*_{NR,n}-M^*_{NR,p})/(M\alpha)=0.187$ by the RBHF theory in the full Dirac space.
The obtained slope is not changed if more data with a step-size $\Delta\alpha = 0.05$ are used.


\begin{figure}[htbp]
  \centering
  \includegraphics[width=8.0cm]{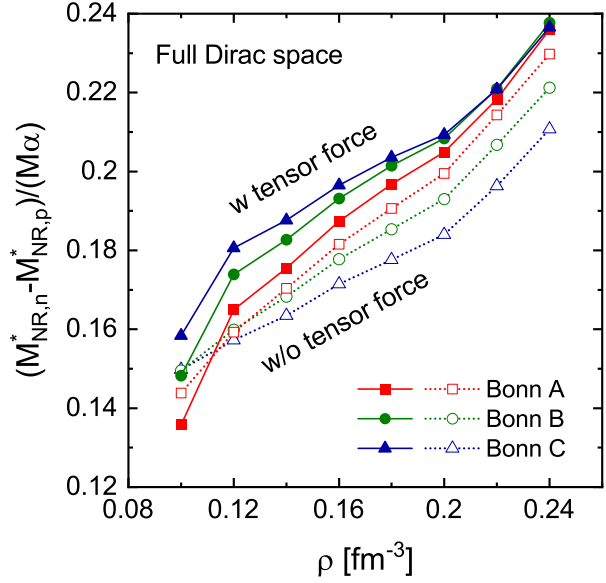}
  \caption{(Color online) The neutron-proton effective mass splitting as a function of the density $\rho$ calculated by the RBHF theory in the full Dirac space with the $NN$ interactions Bonn A, B, and C \cite{Machleidt1989_ANP19-189}.
  The solid lines with solid symbols denote the full calculation with tensor force, while the short-dotted lines with empty symbols denote the self-consistent calculations without tensor force.}
  \label{label-Fig3}
\end{figure}

In Fig.~\ref{label-Fig3}, the density dependence of the magnitude of the neutron-proton effective mass splitting in the full Dirac space is depicted as solid curves with solid symbols by using the same method as shown in Fig.~\ref{label-fig2}.
As for the $NN$ interactions Bonn A, the effective mass splitting increases from $0.136$ at $\rho=0.10\ \text{fm}^{-3}$ to $0.236$ at $\rho=0.24\ \text{fm}^{-3}$.
A similar density dependence of the effective mass splitting is also obtained by using the potentials Bonn B and C \cite{Machleidt1989_ANP19-189}.
At the saturation density $\rho=0.16\ \text{fm}^{-3}$, the magnitudes of the effective mass splitting are 0.187, 0.193, and 0.197 for Bonn A, B, and C, respectively, with difference less than 5\% of the averaged value.
This indicates that the uncertainties of the effective mass splitting from variations of realistic $NN$ interactions are relatively small.


To illustrate the tensor-force effects on the effective mass splitting, we perform the RBHF calculations without the tensor force, i.e., the tensor force from the bare interaction $V$ is removed with the formula derived in Ref.~\cite{2018-WangZH-PhysRevC.98.034313}.
The results are shown in Fig.~\ref{label-Fig3} as short-dotted curves with empty symbols.
It can be seen that the inclusion of the tensor force can increase the effective mass splitting. 
The influence is particularly pronounced in the case of Bonn C due to its strongest tensor force among the Bonn potentials~\cite{Brockmann1990_PRC42-1965}. 
After removing the tensor force, the differences among Bonn potentials still exist. 
This indicates that the central forces also have an influence on the effective mass splitting.


\begin{figure}[htbp]
  \centering
  \includegraphics[width=14.0cm]{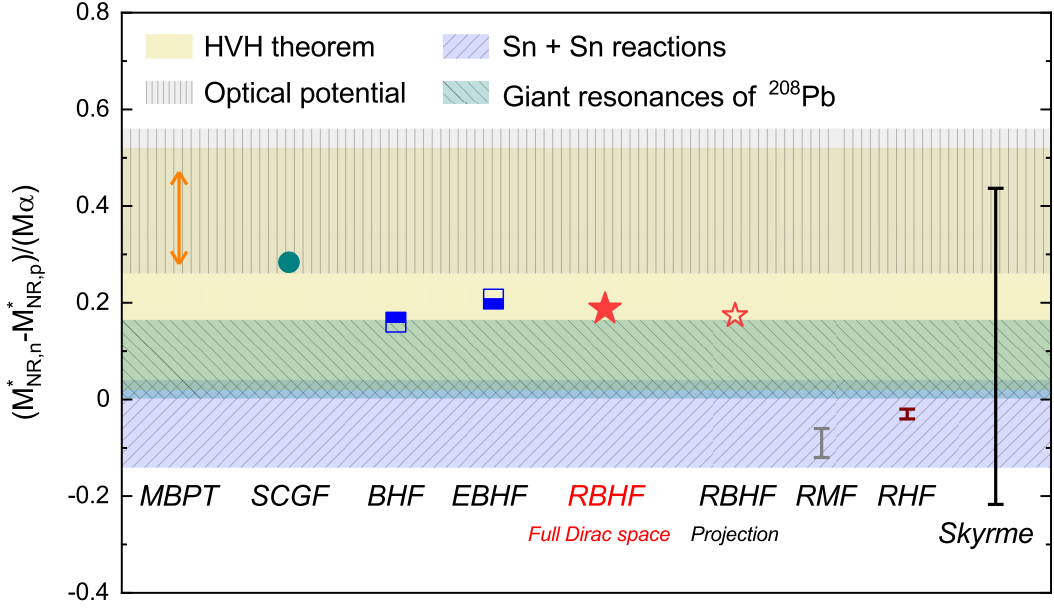}
  \caption{(Color online) The neutron-proton effective mass splitting at the density $\rho=0.16\ \text{fm}^{-3}$ calculated by the RBHF theory in the full Dirac space (red solid star), in comparison with the constraints from analyzing nuclear structure and reaction experiments~\cite{LiXH-2015-Phys.Lett.B743.408, LiBA-2013-Phys.Lett.B727.276, 2020-XuJ-PhysRevC.102.024306, 2019-Morfouace-Phys.Lett.B799.135045} (shaded regions), and the theoretical results obtained by the RBHF theory with the projection method (red empty star), the nonrelativistic BHF~\cite{2021-ShangXL-PhysRevC.103.034316} and extended BHF theory~\cite{2005-Zuo-PhysRevC.72.014005} (blue semifilled squares), the SCGF approach~\cite{2010-Mansour-PTP} (cyan dot), the MBPT calculation with chiral effective forces~\cite{Whitehead-2021-PhysRevLett.127.182502} (orange arrows), the Skyrme-Hartree-Fock theory (black error bar), and the covariant density functional theory~\cite{Margueron-2018-PhysRevC.97.025805} (gray and purple error bars).}
  \label{label-Fig4}
\end{figure}

Finally, we compare the neutron-proton effective mass splitting obtained in this work to the constraints from other experimental analyses as shown in Fig.~\ref{label-Fig4}.
The recent optical potential model analyses of nucleon-nucleus scattering data constrain the effective mass splitting in the range of $(0.41\pm0.15)\alpha$ \cite{LiXH-2015-Phys.Lett.B743.408}.
The analyses based on the HVH theorem using the systematics of nuclear symmetry energy and its slope lead to results $(0.27\pm0.25)\alpha$ \cite{LiBA-2013-Phys.Lett.B727.276}.
Smaller effective mass splitting $(0.083^{+0.081}_{-0.080})\alpha$ at 68\% confidence level is obtained from the Bayesian model analyses of the isoscalar and isovector giant resonances of $^{208}$Pb \cite{2020-XuJ-PhysRevC.102.024306}.
These three experimental analyses give positive values, while a negative value is favored $(-0.05\pm0.09)\alpha$ from a Bayesian analysis of the neutron/proton spectrum ratios in several $\text{Sn}+\text{Sn}$ reactions by using an improved quantum molecular dynamics transport model with Skyrme forces \cite{2019-Morfouace-Phys.Lett.B799.135045}.
It can be found that the values extracted from the experimental observations have relatively large uncertainties and model dependence.
The prediction of the RBHF calculations in the full Dirac space agrees with the analyses based on the HVH theorem and lies between the results obtained from the nucleon-nucleus scattering and the giant resonances of $^{208}$Pb.

Since the experimental constraints on the effective mass splitting have large uncertainties, the \emph{ab initio} calculations can provide important predictions on the neutron-proton effective mass splittings.
In Fig.~\ref{label-Fig4}, available predictions from several \emph{ab initio} methods are listed.
The RBHF theory with the projection method gives the value of $0.17\alpha$.
The effective mass splittings from the BHF theory at $\rho=0.16\ \text{fm}^{-3}$ \cite{2021-ShangXL-PhysRevC.103.034316} and the extend BHF (EBHF) theory at $\rho=0.2\ \text{fm}^{-3}$ \cite{2005-Zuo-PhysRevC.72.014005} are $0.16\alpha$ and $0.21\alpha$, respectively.
It is seen that all the Brueckner calculations give very close predictions.
The SCGF approach by using the CD-Bonn potential suggests the value of $0.28\alpha$~\cite{2010-Mansour-PTP}, which is compatible with the results of the Brueckner calculations.
The MBPT using the chiral forces gives larger predictions in the range $(0.28$--$0.47)\alpha$ \cite{Whitehead-2021-PhysRevLett.127.182502}, where the uncertainty comes from the different choices of chiral potentials.
In general, the effective mass splittings given by the \emph{ab initio} methods have a positive sign and are larger than $0.16\alpha$.

In Fig.~\ref{label-Fig4}, we also show the calculated values for neutron-proton effective mass splitting obtained from several nonrelativistic Skyrme functionals
as well as the covariant density functionals including the relativistic mean field (RMF) theory and the relativistic Hartree-Fock (RHF) theory.
The mean values as well as the standard deviations from RMF theory and RHF theory are $(-0.09\pm0.03)\alpha$ and $(-0.03\pm0.01)\alpha$ \cite{Margueron-2018-PhysRevC.97.025805}, respectively.
It is interesting to find that all the splittings from covariant density functionals are negative \cite{Margueron-2018-PhysRevC.97.025805}.
For RMF theory, where single-particle potentials are independent of momentum/energy, this pattern can be understood in a way that $M^*_{NR,n}-M^*_{NR,p} = -(U_{0,n}-U_{0,p})$ \cite{ChenLW-2007-PhysRevC.76.054316}.
In neutron-rich matter, RMF theories generally have $U_{0,n}>U_{0,p}$ which leads to the isospin splitting $M^*_{NR,n}-M^*_{NR,p}<0$.
Exception exists for a nonlinear point-coupling RMF parameter set with higher-order isovector-vector term like PC-F3~\cite{ChenLW-2007-PhysRevC.76.054316,2008-LiBA-Phys.Rep.464.113}, where $U_{0,n}<U_{0,p}$ is found for density higher than 0.21 fm$^{-3}$.
No definitive conclusion on the sign of $(M^*_{NR,n}-M^*_{NR,p})/M$ can be drawn from the results $(0.127\pm0.310)\alpha$ of Skyrme functionals \cite{Margueron-2018-PhysRevC.97.025805}.
This is an indication that these values are weakly constrained by their fitting protocols, which are mostly based on masses and charge radii of finite nuclei and therefore the isospin properties of most of these functionals are not very well determined. The exception is the Lyon forces, which have been adjusted in particular to properties of neutron-rich nuclei \cite{Chabanat1997_NPA627-710,Chabanat1998_NPA635-231}. In contrast to many of the other Skyrme forces they have a negative effective mass splitting $M^*_{NR,n}-M^*_{NR,p} < 0$ \cite{Lesinski2006_PRC74-044315,2011-Ou-Phys.Lett.B,Margueron-2018-PhysRevC.97.025805}, which is in agreement with the pure mean-field RMF results, but in disagreement with the microscopic RBHF results discussed before. This is an open problem which requires further investigations. Perhaps it can be understood by the fact the single-particle level density at the Fermi surface is much too low for the pure mean-field functionals and that the inclusion of additional correlations by particle vibrational coupling \cite{Litvinova2006_PRC73-044328,Afanasjev2015_PRC92-044317} and by the inclusion of a tensor forces \cite{Otsuka2005_PRL95-232502} can improve this situation.
The predictions from \emph{ab initio} methods will certainly be helpful to constrain the isovector parameters in both nonrelativistic and covariant density functionals.


In summary, the nonrelativistic nucleon effective masses in nuclear matter have been studied with the relativistic Brueckner-Hartree-Fock (RBHF) theory in the full Dirac space.
For symmetric nuclear matter, the neutron and proton effective masses are 0.80, which agrees well with the empirical values $0.8\pm0.1$.
For neutron-rich matter, the effective mass at the Fermi momentum for neutron is larger than that of proton.
By fitting the effective mass splitting with a linear function of the isospin asymmetry, the neutron-proton effective mass splitting at empirical saturation density is obtained as $(M^*_{NR,n}-M^*_{NR,p})/M = 0.187\alpha$.
The density dependence for the effective mass splitting is also studied from 0.1 to 0.24 fm$^{-3}$, and the effective mass splittings are increased by 0.10, 0.09, and 0.08 for potential Bonn A, B, and C, respectively.
By removing the tensor force from the bare interaction, a reduction of the effective mass splitting is found.
The positive value of the neutron-proton effective mass splitting from the RBHF theory in the full Dirac space is quantitatively consistent with the constraints from the nucleon-nucleus scattering, the giant resonances of $^{208}$Pb, and the Hugenholtz--Van Hove theorem with systematics of nuclear symmetry energy and its slope.
The predictions from the RBHF theory in the full Dirac space might be helpful to constrain the isovector parameters in phenomenological density functionals.

\begin{acknowledgments}

S.W. thanks X.-L. Shang and J. Xu for helpful discussions.
This work was supported in part by the National Key R\&D Program of China under Contracts No. 2017YFE0116700 and 2018YFA0404400, the National Natural Science Foundation of China (NSFC) under Grants No. 12205030, 11935003, 11975031, 11875075, 12070131001, and 12047564, the Fundamental Research Funds for the Central Universities under Grants No. 2020CDJQY-Z003 and 2021CDJZYJH-003, the MOST-RIKEN Joint Project "Ab initio investigation in nuclear physics",
the Deutsche Forschungsgemeinschaft (DFG, German Research Foundation) under Germany’s Excellence Strategy EXC-2094-390783311, ORIGINS, and the Institute for Basic Science under Grant No. IBS-R031-D1.
Part of this work was achieved by using the supercomputer OCTOPUS at the Cybermedia Center, Osaka University under the support of Research Center for Nuclear Physics of Osaka University and the High Performance Computing Resources in the Research Solution Center, Institute for Basic Science.

\end{acknowledgments}

\bibliography{EffectiveMass-short}

\end{document}